\documentclass[journal]{IEEEtran}

\usepackage{enumitem}
\usepackage{amsmath}
\usepackage{url}
\usepackage[labelfont=bf]{caption}
\usepackage{subcaption}
\usepackage{booktabs, array, multirow}
\usepackage[algoruled]{algorithm2e}
\usepackage{algorithmic}
\usepackage{listings}
\usepackage{fancyvrb}
\usepackage
[acronym,smallcaps,nowarn,section,nogroupskip,nonumberlist]{glossaries}
\glsdisablehyper{}
\usepackage{threeparttable}
\usepackage{float}
\usepackage{graphicx}
\usepackage{tabularx}
\usepackage{booktabs} 
\usepackage{amsopn}
\usepackage{amsfonts}
\usepackage[numbers,sort&compress]{natbib}
\usepackage[dvipsnames]{xcolor}
\definecolor{c1}{HTML}{ce2626}
\usepackage{hyperref}
\usepackage[nameinlink]{cleveref}

\ifCLASSINFOpdf
\else
\fi
\begin{document}
%
\title{Cross-modal Variational Auto-encoder for Content-based Micro-video Background Music Recommendation}
%
%
%

\author{Jing~Yi, Yaochen~Zhu, Jiayi~Xie,
        Zhenzhong~Chen,~\IEEEmembership{Senior Member,~IEEE}
        
\thanks{This work was supported in part by the National Natural Science Foundation of China (Grant No. 62036005). }

\thanks{J. Yi and Z. Chen are with the School of Computer Science, Wuhan University, Hubei 430079, China. Y. Zhu, J. Xie and Z. Chen are with the School of Remote Sensing and Information Engineering, Wuhan University, Hubei 430079, China. (Corresponding author: Zhenzhong Chen, zzchen@whu.edu.cn)}
}

\markboth{IEEE TRANSACTIONS ON MULTIMEDIA,~VOL.~22, NO.~12, July~2021}%
{Shell \MakeLowercase{\textit{et al.}}: Bare Demo of IEEEtran.cls for IEEE Journals}

\maketitle

\begin{abstract}
In this paper, we propose a cross-modal variational auto-encoder (CMVAE) for content-based micro-video background music recommendation. CMVAE is a hierarchical Bayesian generative model that matches relevant background music to a micro-video by projecting these two multimodal inputs into a shared low-dimensional latent space, where the alignment of two corresponding embeddings of a matched video-music pair is achieved by cross-generation. Moreover, the multimodal information is fused by the product-of-experts (PoE) principle, where the semantic information in visual and textual modalities of the micro-video are weighted according to their variance estimations such that the modality with a lower noise level is given more weights.
Therefore, the micro-video latent variables contain less irrelevant information that results in a more robust model generalization. Furthermore, we establish a large-scale content-based micro-video background music recommendation dataset, TT-150k, based on approximately 3,000 different background music clips associated to 150,000 micro-videos from different users. Extensive experiments on the established TT-150k dataset demonstrate the effectiveness of the proposed method. A qualitative assessment of CMVAE by visualizing some recommendation results is also included.
\end{abstract}


%
\IEEEpeerreviewmaketitle

\section{INTRODUCTION}
\label{section1}

Nowadays, micro-videos have become an increasingly prevalent medium on the Web. Compared with texts or pictures, micro-videos contain rich visual contents, audio sounds, as well as textual tags and descriptions, which allow people to more vividly record and share their daily lives. {Applications such as TikTok\footnote{\url{https://www.tiktok.com/}} and Kwai\footnote{\url{https://www.kuaishou.com/}} have witnessed the success of the micro-video sharing platform with a huge user base. 
Matching micro-videos with suitable background music can help uploaders better convey their contents and emotions. Moreover, background music selection is one of the key elements to make the generated videos attractive and increase the click-through rate of the uploaded videos.} However, manually selecting the background music becomes a painstaking task due to the voluminous and ever-growing pool of candidate music. Therefore, {automatically recommending background
music to videos could help Online micro-video sharing platforms attract more users and spread the micro-videos, which is a crucial task for multimedia cross-modal matching.}

Existing cross-modal matching methods have mainly focused on the matching between textual and visual (image or video) modalities. For example, Xu \textit{et al.} \cite{xu2015jointly} proposed a joint video-language framework for text-based video retrieval evaluated on \cite{chen2011collecting}. Semedo and Magalhães \cite{mm2020_temporal} proposed a new temporal constraint for image-text matching using the NUS-WIDE benchmark \cite{NUS-WIDE}. Zhang \textit{et al.} \cite{zhang2020context} utilized inter-modal attention to discover semantic alignments between words and image patches for image-text retrieval, where the intra-modal attention {was} also exploited to learn semantic correlations of fragments for each modality. Meanwhile, metric-learning-based methods have been proposed to better implement cross-modal matching \cite{9178501,7801952,wei2020universal}. The continual success in such areas is facilitated by the establishment of Flickr \cite{plummer2015flickr30k}, MS-COCO \cite{lin2014microsoftcoco}, MSR-VTT \cite{xu2016msr}, \textit{etc.}, which are widely employed benchmarks for the evaluation of visual-textual retrieval models. He \textit{et al.} \cite{mm2019_newdataset} further proposed a new dataset for fine-grained cross-modal retrieval. However, there exists no publicly available dataset for matching videos with background music, which becomes the main obstacle for future research regarding the automatic recommendation of background music to videos. 

It is very challenging to construct such a video background music recommendation dataset. For example, how to collect different music clips and acquire different videos using the music clips as background music is a problem. Taking the issues into consideration, we manage to establish a micro-video background music matching dataset, which we name TT-150k, based on the popular micro-video sharing platform TikTok. Specifically, the candidate music clips are selected from the TikTok pop charts and the background music adopted by popular micro-videos. The established TT-150k dataset includes extracted features from more than 3,000 music clips and about 150,000 micro-videos that use the music clips from the candidate list. In our establishment, we ensure that the popularity of the music clips is proportional to its true distribution in the population of TikTok, and therefore the dataset can faithfully reflect the music popularity distribution in the real-world scenario. 

With the TT-150k dataset established, we aim to design an effective algorithm for the automatic matching of micro-videos with background music. Previous studies on video-music matching \cite{liu2018background,li2019querybyvideo} mainly rely on manually annotated emotion tags to match music and videos in the affective space. However, such manual annotation is laborious and time-consuming for large datasets. Therefore, it is imperative to seek methods that match music to videos without manually-labeled information. Moreover, the semantic structure of micro-videos can vary drastically with that of the music: A silent micro-video is a hodgepodge of visual and textual information, while the music contains only audio information. Therefore, it is challenging to match the semantic-rich video latent space with the monotonous music latent space. 
{In addition, the weak association between micro-videos and background music makes it hard to learn the matching patterns with the typical matching loss \cite{wang2018learning}. }
To address the above challenges, we propose a {C}ross-{M}odal {V}ariational {A}uto-{E}ncoder (CMVAE) for content-based micro-video background music recommendation. CMVAE is a hierarchical Bayesian generative model that matches relevant background music to micro-videos by projecting them into a shared low-dimensional latent space. Specifically, the latent space is constrained by cross-generation such that the embeddings of a matched video-music pair are closer than that of an unmatched pair. Through this, a more accurate similarity measurement between micro-videos and music can be obtained and utilized for the recommendation. Meanwhile, to fully utilize the heterogeneous visual and textual information of the micro-video for matching, the bimodal information in micro-videos is comprehensively fused according to the product-of-experts (PoE) principle. In this way, the information in each modality is weighted by the reciprocal of its variance estimation to give the modality with a lower noise level more weights, such that irrelevant information in the micro-video embeddings can be reduced. Besides, the PoE fusion is shown to be robust to the textual modality missing problem, which is a commonly encountered problem when performing micro-video analysis. The main contributions of this paper can be concretely summarized as follows:

\begin{itemize}[]
    \item We propose a novel hierarchical Bayesian generative model, CMVAE, for content-based micro-video background music recommendation. {CMVAE projects micro-video and background music into a shared low-dimensional latent space, where the alignment of two corresponding embeddings of a matched video-music pair is achieved by constraining their latent variables to generate each other. Through the cross-generation, a better alignment of the latent pairwise distributions and matching between the micro-video and music can be realized in the latent space for recommendations.}
    \item The product-of-experts (PoE) principle is adopted to fuse the modality-specific embeddings of visual and textual modalities of the micro-video so that the complementary multimodal information can be better exploited. With PoE, the semantic information in different modalities are weighted according to their variance estimations for giving more weights to the modality with a lower noise level, such that the micro-video embeddings could contain less irrelevant information for a more robust generalization. 
    \item A large-scale content-based micro-video background music recommendation dataset, TT-150k, is established. The dataset contains extracted features from more than 3,000 candidate music clips and about 150k corresponding micro-videos, where the popularity distribution is consistent with that of the real-world scenario. On the established TT-150k dataset, experiments show that the proposed CMVAE significantly outperforms the state-of-the-art methods.

\end{itemize}

The remainder of this article is organized as follows. Section \ref{section2} gives a literature review of the work related to our proposed method. Section \ref{section3} introduces the establishment of the dataset TT-150k. In Section \ref{section4}, we illustrate the proposed CMVAE with details. Section \ref{section5} presents the experimental settings and analyzes the experimental results. Finally, Section \ref{section6} briefly summarizes the article.

\section{RELATED WORK} \label{section2}
\subsection{Audiovisual Cross-modal Matching}

Cross-modal matching aims to match relevant materials where the modality of the target is different from that of the query \cite{cross_modal_review}.
Compared with the matching of two sources that come from a single modality, cross-modal matching requires measuring the similarity of heterogeneous sources composed of different modalities, and therefore it is a more challenging task. Recent researches have mainly focused on matching between textual and visual modalities \cite{mm2019_cross,mm2019_annotaton,zhen2019deep,wei2020universal,image_text,8931635}, using open-sourced datasets such as MSCOCO \cite{lin2014microsoftcoco} and Flickr \cite{plummer2015flickr30k}. { The need for matching audio and visual contents, such as music video generation, has existed for a long time with many efforts dedicated to solving the problem \cite{4130378,10.1145/2964284.2967245,4217514}. }

{Music recommendations have existed for a long time with much attention \cite{NIPS2013_b3ba8f1b,10.3389/fams.2019.00044}.  For example, Andjelkovic \textit{et al.} \cite{ANDJELKOVIC2019142} integrated content and mood-based filtering which utilized a music-specific model of affect, rather than the traditional Circumplex model to explore music mood. Vystr\v{c}ilov\'{a} and Pe\v{s}ka \cite{lyrics_music_rec} emphasized the importance of lyrics-based embedding for music recommendations. Cheng \textit{et al.} \cite{emps} explored the effects of music play sequence on developing effective personalized music recommender systems using word embedding techniques.
Cheng and Shen \cite{location_music_rec} presented a novel venue-aware music recommender system to effectively identify suitable songs for various types of popular venues in our daily lives, where songs and venue types were represented in the shared latent space and can be directly matched.}

{However, background music recommendation for visual contents has been less explored.
Several pioneering methods for audiovisual cross-modal matching }include: Chao \textit{et al.} \cite{chao2011tunesensor} recommended background music for digital photo albums using the relatedness between image tags and music mood tags. Li and Kumar \cite{li2019querybyvideo} utilized emotion tags as metadata to align the music and video modalities. These emotion tags for the video, however, were manually annotated from crowdsourcing, which is high in labor and time costs. Chen \textit{et al.} \cite{liu2018background} relied on Thayer's emotion model \cite{thayer1990biopsychology} and the videos and audios falling in the same quadrant of emotions were regarded as positive samples. 
{Shah \textit{et al.} \cite{advisor} proposed a personalized video soundtrack recommendation system by matching mood tags of videos and music with mood clusters as ground truth, where the mood tags of videos were extracted from location information with sensor-annotated data and video content. Sasaki \textit{et al.} \cite{DBLP:conf/mmm/SasakiHOM15}  matched an input video with music using an emotional plane through the valence/arousal model of affect \cite{mood}. Shin and Lee \cite{7881714} also utilized the emotion (arousal and valence) for video and music pair where the music was arranged with the video using emotion similarity. Shang \textit{et al.} \cite{9381371} aimed to automatically retrieve relevant music on social media based on the connotation that implicitly expressed the abstract idea or inherent emotion beyond the explicit content of visual inputs. 
A metaphor-enriched connotation extraction module  was exploited to explicitly identify metaphors through a set of semantic and emotion
entities extracted from both the image and music using crowdsourcing image-to-song labels.
}
Lin \textit{et al.} \cite{lin2014semantic} argued the importance of matching the rhythm of the music with the visual movement of the video, where the tags of videos provided by users were used to calculate their relationship with song lyrics of music candidates.

{Above methods heavily depend on the emotion labels to match background music for visual contents. Wu \textit{et al.} \cite{mm12}, however, proposed content-based cross-modal matching of music and images, where features are extracted first and ranking canonical correlation analysis (CCA) was utilized to model the local relationship between video and music for each cluster.}
Suris \textit{et al.} \cite{suris2018cross} further employed the visual features and audio features provided by Youtube-8M \cite{abu2016youtube-8m} to constrain the visual and audio embeddings of the same video as close as possible and {predicted} the corresponding label of the video. Wu \textit{et al.} \cite{music_video} constructed a music-image dataset with manual annotation to explore the matching patterns between the two modalities using CCA-based method. CBVMR \cite{hong2018cbvmr} introduced a content-based retrieval model that only used the matching signal between music and videos without any metadata like emotions. However, since there exists no publicly available dataset, these methods are not directly comparable to each other, which motivates us to establish the content-based micro-video background music recommendation dataset, TT-150k, in this paper. 
{Moreover, manually annotating emotion tags to match music and videos in the affective space is laborious and time-consuming for large datasets, and therefore, we seek for content-based methods without any additional annotation. However, since the matching status between video-music pairs is extracted from micro-videos which are User-generated content (UGC), the professionalism of the matching might be worse than crowdsourcing. Considering this, we propose the cross-generation module to overcome the weak supervision of the matching status.}

\subsection{Variational-based Recommendation}
{Variational auto-encoder (VAE) provides a probabilistic manner for describing an observation in the latent space which takes data as input $\mathbf{x}$ and discovers latent variable $\mathbf{z}$ with a probability distribution to govern the generation of $\mathbf{x}$. VAE aims to find an approximation to the intractable posterior $p(\mathbf{z} \mid \mathbf{x})$ of $\mathbf{z}$ from the distribution family $q_{\phi}(\mathbf{z} \mid \mathbf{x})$. Typically, parameters of the variational posterior distribution are obtained through an encoder network, where the likelihood conditioned on the latent variable $\mathbf{z}$ is generated from a decoder network.}
The objective of VAE is to maximize the marginal log-likelihood $\log p{\mathbf{(x)}}$, which is proved to be equivalent to optimizing an Evidence Lower BOund (ELBO) \cite{kingma2013auto}:
\begin{equation}
\begin{aligned}
    &\mathcal{L}\left(\mathbf{x}^{(i)}; \theta, \phi\right)= \\
    &\mathbb{E}_{\mathbf{z} \sim q_{\phi}}\left[\log p_{\theta}\left(\mathbf{x}^{(i)} \mid \mathbf{z}\right)\right]-D_{\text{KL}}\left(q_{\phi}\left(\mathbf{z} \mid \mathbf{x}^{(i)}\right) \mid p(\mathbf{z})\right),
    \label{vae}
\end{aligned}
\end{equation}

\noindent where $\mathbf{x}^{(i)}$ is sampled from the observation population. $\theta$ and $\phi$ are parameters to be trained of the decoder and encoder, respectively.

Compared to auto-encoder (AE) \cite{ae}, which would suffer from performance degradation when facing noisy inputs, and denoising auto-encoder (DAE) \cite{dae}, which adds fixed noise to the input during the training phase and cannot handle diverse noise for different samples, VAE could capture semantic structures of high-dimensional data into latent variables with the inferred latent embeddings corrupted with dynamic Gaussian noise to improve the robustness. Owing to its advantages, VAE has recently been extended for recommendations. For example,
{MultiVAE \cite{liang2018variational} assumed a multinomial likelihood of the click feedback of a user with a variational auto-encoder to model user's interactions among items.}
MacridVAE \cite{ma2019learning} extended MultiVAE by constraining the user representations to be disentangled in both the macro and the micro-level. On the other hand,
{Li and She \cite{li2017collaborative} utilized VAE to model item contents where the item content VAE was coupled into the probabilistic matrix factorization model. In this way, item contents and collaborative information are fully considered for better recommendations. Wang \textit{et al.} \cite{website_vae} proposed a side information-aided website recommendation system that used the browsing history of a set of users and their side information to predict the recommended websites. Both user-website interaction information and side information were treated as input, and a VAE model was adopted to generate user's interested websites from partial observations.
Chen and de Rijke \cite{10.1145/3270323.3270326} proposed to simultaneously recover user ratings and side information of items by using a VAE, where user ratings and side information were encoded and decoded collectively through the same inference network and generation network. Due to the heterogeneity of user ratings and side information, the final layer of the generation network followed different distributions. Karamanolakis \textit{et al.} \cite{vae_prior} exploited online item reviews to collaborative filtering methods by replacing the user-agnostic standard Gaussian prior with heterogenous, user-dependent priors which were estimated empirically as functions of the user’s review text. Yi and Chen \cite{mvgae} proposed a multimodal variational graph auto-encoder method that learned a Gaussian variable for each node and the modality-specific Gaussian node embeddings were fused according to the product-of-experts principle such that the semantic information in each modality was weighted based on the impotence level to the recommendation.}
However, the utilization of VAE for multimodal cross-modal retrieval is comparatively less explored, which leads us to propose CMVAE for micro-video background music recommendation in this work.

\section{Introduction of The TT-150k dataset} \label{section3}

To bridge the gap that there exists no publicly available matching dataset that associates videos with background music, we establish TT-150k to consider the practical scenario where a music clip could be associated to multiple videos while the relevant datasets such as Youtube-8m \cite{abu2016youtube-8m} mainly associate one music clip to one video only. TT-150k is intended to be a benchmark for micro-video background music recommendation. It is collected on the popular micro-video sharing platform TikTok, where numerous videos are uploaded every day with fantastic background music. The central criterion for establishing this dataset is to faithfully reflect the distribution of micro-videos and candidate music clips in the real-world scenario to facilitate research regarding the discovery of matching patterns between music and micro-videos. Then, relevant background music can be automatically recommended upon the upload of a micro-video.

\subsection{Dataset Introduction}

We first built a music candidate list based on approximately 3,000 music clips. A music clip was selected into the list if it satisfied one of the two criteria: 1) it appeared on the TikTok pop charts, or 2) it was used in a randomly collected popular video (which ensures the quality of the background music). Note that the background music clip of a popular micro-video is not necessarily a popular music clip, and therefore the popularity distribution of the music clips in the candidate list can still faithfully reflect the real-world scenario. The music clips with the title 
``original music" were eliminated because these music clips are often made and uploaded by users that are tailored for a certain video, which may not be able to generalize well to other micro-videos. Besides, varied remixes of the same music {were} indexed as different music clips in our dataset. With the music candidate list established, we gathered a list of videos that use a certain music clip in the above music set as background music according to TikTok's song search function. We also collected the number of videos using a certain music clip as background music (\textit{i.e.}, music usage amount) at the same time. 

\begin{figure}
\centering
\includegraphics[scale=0.85]{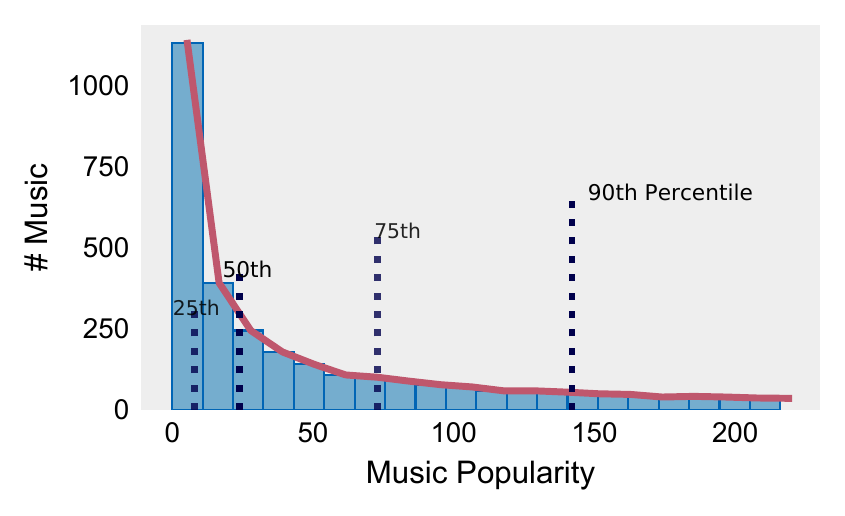}
\caption{Popularity distribution of the candidate music clips in our TT-150k dataset.}
\label{FIG:sampling}
\end{figure}

To approximate the true distribution of music adoption, \textit{i.e.,} its popularity, we used the collected music usages to sample videos with the principle to make the relative popularity of these music clips as close as possible to the relative popularity of the music clips in the real-world scenario. Specifically, we used a Gamma distribution to fit the distribution of music usage amount through Maximum Likelihood Estimation. The specific popularity distribution of the candidate music clips after sampling is shown in Figure \ref{FIG:sampling}. With the sampling strategy, we can reduce the number of videos without losing the original distribution of real-world data, which makes it suitable to analyze the matching pattern of videos and background music. After calculating the number of videos that needed to be gathered based on the relative music popularity distribution, we sampled the latest uploaded videos for different music clips. At the same time, brief descriptions or several hashtags attached to the video were also collected if they exist. In this way, the micro-videos consist of bimodal representations of visual and the corresponding textual contents. 

\subsection{Feature Engineering}

For visual information of the micro-videos, we preprocess the videos and extract video-level features using an effective pre-trained ResNet model \cite{he2016deep}. The structure of ResNet enhances the quality of image representations by refining the raw information from the input images in a cascade manner by residual modeling. Specifically, 
we first utilize FFmpeg\footnote{\url{https://www.ffmpeg.org/}} to extract video frames at three frames per second. Then we could use the pre-trained model such as ResNet \cite{he2016deep} or I3D \cite{i3d} to extract the visual features. We adopt ResNet in this work for its simplicity and follow the work \cite{abu2016youtube-8m} of temporal global average pooling to fuse features extracted from different frames to get final video-level features. 

For textual information of the micro-videos, observing that TikTok is an international micro-video sharing platform where the languages the users use are diverse, we use the multi-lingual Bert-M model \cite{devlin2018bert} to extract the textual features. Bert-M is pre-trained on a large corpus composed of 104 languages where multi-lingual aligned semantic structures could be learned.  

For music features, we extract the spectrograms from the raw audio clips, and then utilize the Vggish network \cite{hershey2017vggish} pre-trained on AudioSet \cite{audioset} to extract the song-genre-related features. Moreover, we exploit openSMILE \cite{eyben2010opensmile} to extract the pitch and emotion-related features. Specifically, we reduce the dimension of openSMILE features to the same size as Vggish features with principal component analysis (PCA) after normalizing them to zero mean and unit variance.

\subsection{Dataset Statistics}

\begin{figure}
\centering
\includegraphics[scale=0.57]{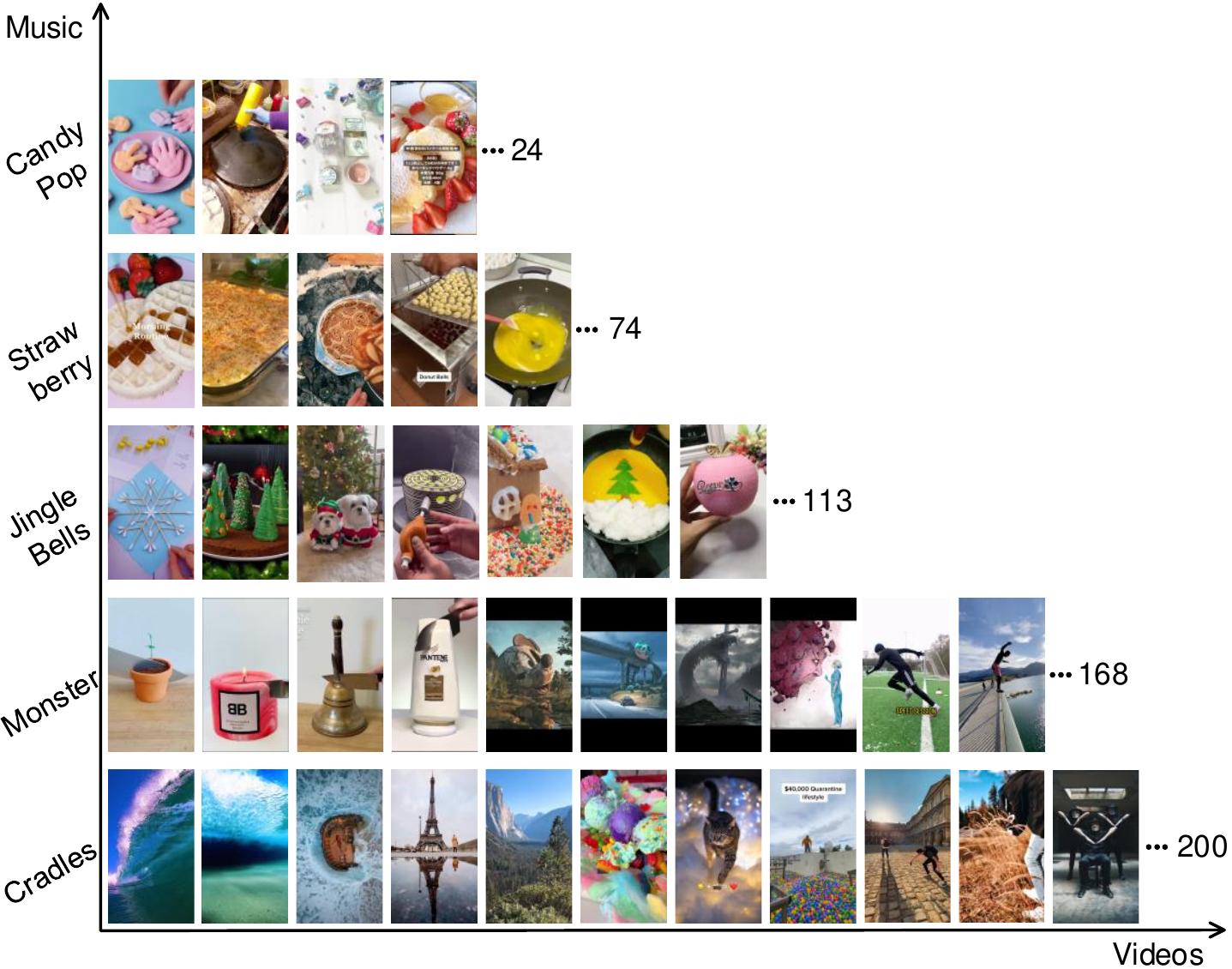}
\caption{An exemplar subset of videos and their matched background music for the established TT-150k dataset.}
\label{FIG:dataset_brief}
\end{figure}

The statistics of the micro-video background music recommendation dataset are presented in Table \ref{TAB:DATA_STATISTIC}. In summary, the established TT-150k dataset contains extracted features from 3,003 music clips and 146,351 videos, where a music clip is selected by as least 3 micro-videos and at most 219 micro-videos. Figure \ref{FIG:dataset_brief} shows an exemplar subset of music clips and micro-videos for the established TT-150k dataset. TT-150k is established based on videos that use a certain music clip in the candidate music set as the background music. In this way, for each video, the ground truth is the background music that the uploader chose for the video.

\begin{table}[htbp]
\centering
\caption{Statistics of the established TT-150k dataset.}
\label{TAB:DATA_STATISTIC}
\begin{tabular}{cccc}
\toprule
\#Music & \#Video  &avg $\pm$ std \#v/\#m & min/max \#v/\#m  \\ 
\midrule
 3,003 &  146,351 &  49  $\pm$ 57 &  3 / 219    \\
\bottomrule
\end{tabular}
\end{table}

\section{METHODOLOGY} \label{section4}

\begin{figure*}
\centering
\includegraphics[scale=0.36]{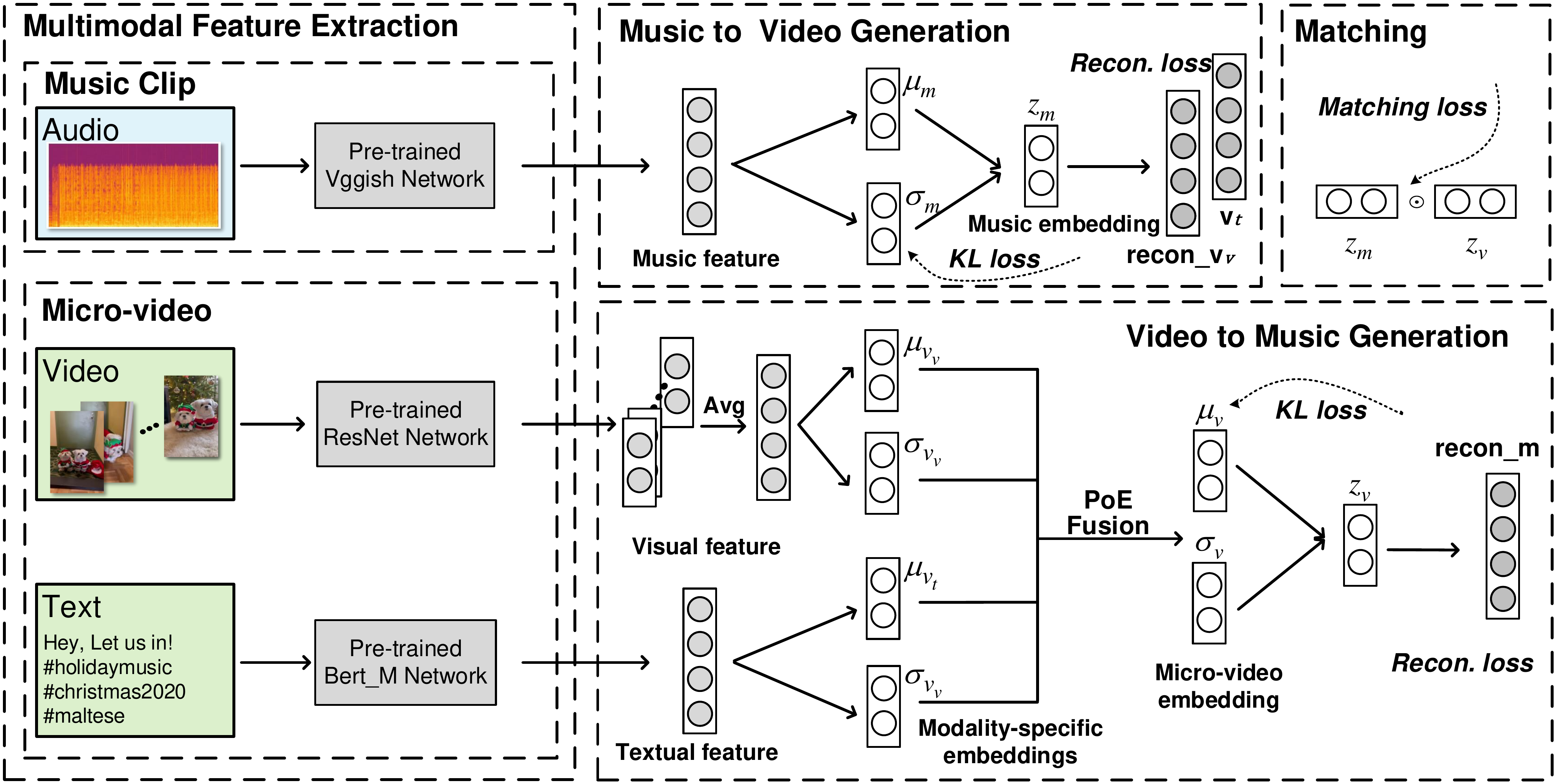}
\caption{The framework of our proposed CMVAE for background music recommendation of micro-videos. Specifically, music feature $m$ is encoded into the latent variable $\mathbf{z}_m$ which follows $\mathcal{N}(\mathbf{\mu}_m, \mathbf{\sigma}_m)$ by an MLP. Video features ($\mathbf{v}_v$ and $\mathbf{v}_t$) are fed into modality-specific encoders with the latent variables from visual and textual modalities fused by the product-of-experts (PoE) principle to compute $\mathbf{z}_v$. The final loss function consists of reconstruction losses, KL divergence losses, and the matching loss.}
\label{FIG:MODEL}
\end{figure*}

The overall framework of CMVAE is shown in Figure \ref{FIG:MODEL}. Specifically, CMVAE aims to align the latent embeddings of music and videos via a cross-modal generative process to model the matching of the music and videos. With the generative process defined, the intractable posterior distributions of the latent variables are estimated by variational inference. Furthermore, to effectively utilize the comprehensive multimodal video contents, CMVAE fuses the modality-specific latent variables by a product-of-experts system, where the information in each modality is weighted by its importance for the matching purpose. The details of the proposed model are expounded in the following sections.
{Notations used in this article are summarized in Table \ref{TAB:notation}.}

\begin{table}[t]
\centering
\caption{{Notations used in our method.}}
\begin{tabular}{ll}
\toprule
  Notation& Description  \\ 
\midrule
$\mathbf{m} \in \mathcal{M}$& a music feature in music set $\mathcal{M}$ \\
$\mathbf{v} \in \mathcal{V}$& a video feature in video set $\mathcal{V}$  \\
$\mathbf{v}_v, \mathbf{v}_t$&  visual and textual features of a video\\
$y$& indicator of the matching status\\
$\mathrm{f}$&  mapping of triads $\{(\mathbf{m}, \mathbf{v}, y)\}$ \\
$d$& the dimension of latent variables\\
$\mathbf{z}_m$&  the music latent variable \\ 
$\mathbf{z}_v$&  the video latent variable\\
$\mathbf{z}_{v_v}, \mathbf{z}_{v_t}$& visual and textual variables of a video\\
$\mathbf{\mu}_m, \mathbf{\sigma}_m$&  mean and variance of the music latent variable  \\ 
$\mathbf{\mu}_v, \mathbf{\sigma}^{2}_v$&  mean and variance of the video latent variable  \\ 
$\mathbf{\mu}_{v_v}, \mathbf{\sigma}^{2}_{v_v}$& mean and variance of the video visual variable\\
$\mathbf{\mu}_{v_t}, \mathbf{\sigma}^{2}_{v_t}$& mean and variance of the video textual variable  \\
$p(\mathbf{z}_m \mid \mathbf{m})$& true posterior of the music latent variable \\
$p(\mathbf{z}_v \mid \mathbf{v})$& true posterior of the video latent variable \\
$q(\mathbf{z}_m \mid \mathbf{m})$& variational posterior of the music latent variable\\
$q(\mathbf{z}_v \mid \mathbf{v})$& variational posterior of the video latent variable\\
$p(\mathbf{m} \mid \mathbf{z}_m)$& conditional likelihood of the music latent variable\\
$p(\mathbf{v} \mid \mathbf{z}_v)$& conditional likelihood of the video latent variable\\
$p( \mathbf{z}_m)$& prior of the music latent variable\\
$p( \mathbf{z}_v)$& prior of the video latent variable\\
$f(\cdot)$& the non-linear function\\
\bottomrule
\end{tabular}
\label{TAB:notation}
\end{table}

\subsection{Problem Definition}

Suppose we have a set of music $\mathcal{M}$ and a set of video $\mathcal{V}$, where {each music $\mathbf{m} \in \mathcal{M}$ is associated with a music feature from its audio clip.} Each video $\mathbf{v} \in \mathcal{V}$ is associated with a visual feature $\mathbf{v}_{v}$ extracted from the image sequence and a textual feature $\mathbf{v}_{t}$ from its description. For each music $\mathbf{m} \in \mathcal{M}$ and video $\mathbf{v} \in \mathcal{V}$, we define a mapping $\mathrm{f}: \mathcal{V} \times \mathcal{M} \rightarrow \{0, 1\}$ that depicts whether or not music $\mathbf{m}$ matches a video $\mathbf{v}$. The mapping $\mathrm{f}$ induces a set of triad $\{(\mathbf{v}, \mathbf{m}, y)\}$, where $y=\mathrm{f}(\mathbf{v}, \mathbf{m})$ is the matching indicator.  Given a new video $\mathbf{v}$, our goal is to retrieve a list of music candidates $\mathcal{C}(\mathbf{m}) \subset \mathcal{M}$ where each music $\mathbf{m} \in \mathcal{C}(\mathbf{m})$ is a potential match for the query video.

\begin{figure}
\centering
\includegraphics[scale=0.22]{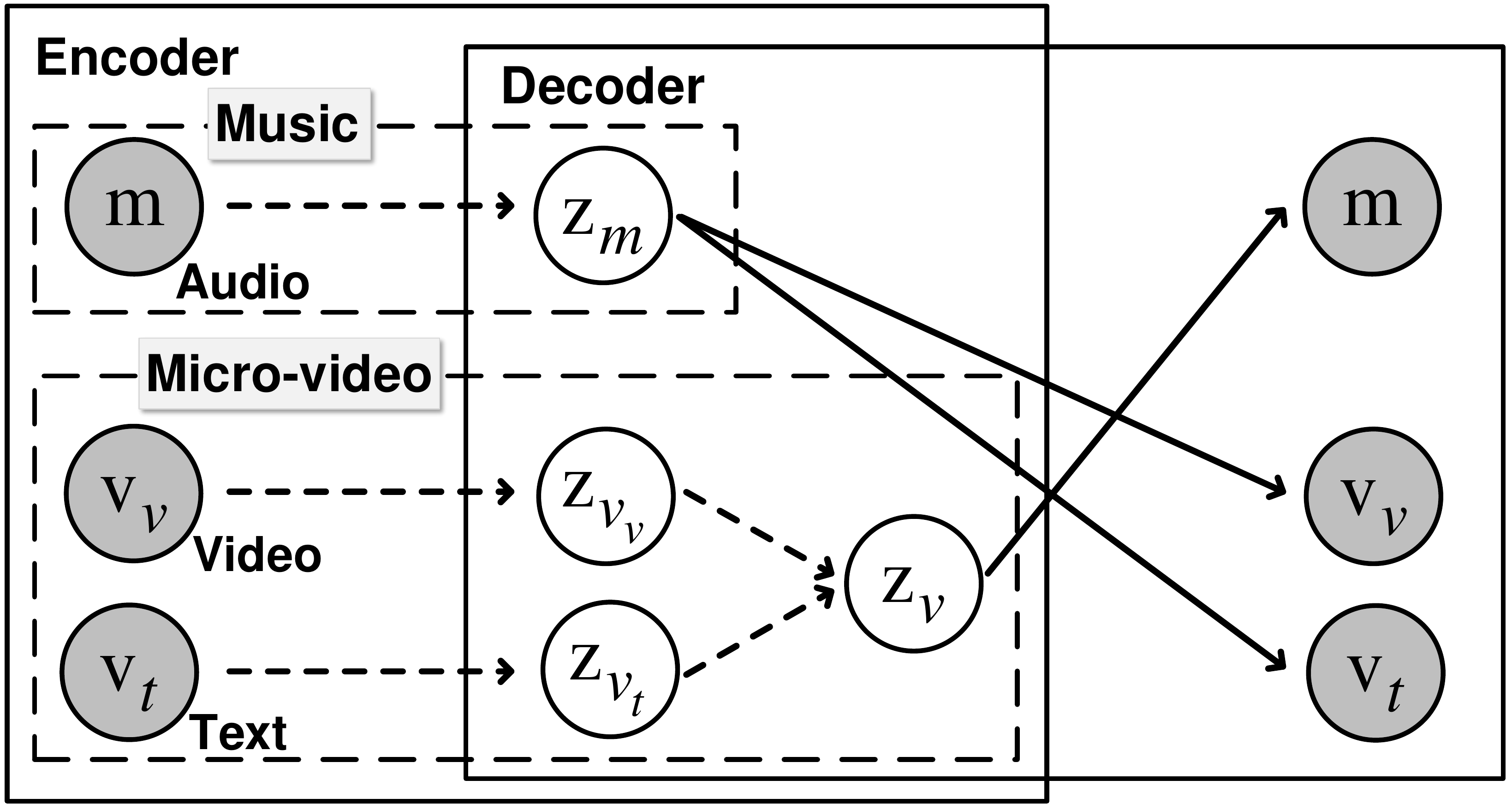}
\caption{The probabilistic graphical model (PGM) of the proposed CMVAE. Hierarchically, we first model modality-level latent embeddings, and then the micro-video-level latent embedding $\mathbf{z}_v$ is obtained for cross-generation.}
\label{FIG:MODEL_brief}
\end{figure}

\subsection{Cross-modal Variational Auto-encoder}
Given a video-music pair $(\mathbf{v}, \mathbf{m})$, considering that their original feature spaces are high-dimensional and misaligned, CMVAE aims to first map $\mathbf{v}$ and $\mathbf{m}$ to a shared $d$-dimensional Gaussian latent space $\mathbb{R}^{d}$ where their matching degree can be properly judged. Such a latent space should have the property that the distance of latent variables for a matched video-music pair is closer than that of an unmatched pair. In CMVAE, this is achieved by constraining the music latent embedding $\mathbf{z}_{m}$ and video latent embedding $\mathbf{z}_{v}$ to be able to generate the video feature $\mathbf{v}$ and music feature $\mathbf{m}$ for a matched pair, \textit{i.e.,} cross-generation. Based on such criteria, the generation process can be formulated as:
\begin{align}
p(\mathbf{v} \mid \mathbf{m})&=\mathbb{E}_{\mathbf{z}_{m} \sim p\left(\mathbf{z}_{m} \mid \mathbf{m}\right)}\left[p\left(\mathbf{v} \mid \mathbf{z}_{m}\right)\right] \label{eq:recon_1} \\
p(\mathbf{m} \mid \mathbf{v})&=\mathbb{E}_{\mathbf{z}_{v} \sim p\left(\mathbf{z}_{v} \mid \mathbf{v}\right)}\left[p\left(\mathbf{m} \mid \mathbf{z}_{v}\right)\right],
\end{align}
{where the process could be viewed as an auto-encoding procedure with cross-generation. Take Eq. (\ref{eq:recon_1}) for an example, the music feature $\mathbf{m}$ is first encoded into the music latent embedding $\mathbf{z}_{m}$, and then the video feature $\mathbf{v}$ is cross generated based on $\mathbf{z}_{m}$. In this way, the alignment of two corresponding embeddings of a matched video-music pair is achieved by constraining their latent variables to generate each other.}
{ Specifically, we first draw the video latent variable $\mathbf{z}_{v}$ and music latent variable $\mathbf{z}_{m}$ in a latent low-dimensional space from their posterior distributions $p\left(\mathbf{z}_{v} \mid \mathbf{v}\right)$ and $p\left(\mathbf{z}_{m} \mid \mathbf{m}\right)$. The features of the video $\mathbf{v}$ and music $\mathbf{m}$ are then generated from the music latent variable $\mathbf{z}_{m}$ and item latent variable $\mathbf{z}_{v}$ through generation neural networks, \textit{e.g.}, MLPs as with variational auto-encoder \cite{kipf2016semi}. 
}

In addition to the cross-generation, we also define the generation of the matching status, \textit{i.e.}, $\mathrm{f}(\mathbf{m}, \mathbf{v})$, by the matching generative distribution $p(y \mid \mathbf{m}, \mathbf{v})$, where the probability depicts the matching degree of the given video-music pair. Since we measure the matching degree of $\mathbf{m}$, $\mathbf{v}$ in the latent space via $\mathbf{z}_{m}$, $\mathbf{z}_{v}$,  $p(y \mid \mathbf{m}, \mathbf{v})$ can be specified as the expected conditional distribution:
\begin{equation}
    p(y \mid \mathbf{m}, \mathbf{v})=\mathbb{E}_{\mathbf{z}_{m} \sim p\left(\mathbf{z}_{m} \mid \mathbf{m}\right), \mathbf{z}_{v} \sim p\left(\mathbf{z}_{v} \mid \mathbf{v}\right)}\left[p\left(y \mid \mathbf{z}_{m}, \mathbf{z}_{v}\right)\right].
\end{equation}
{In this paper, we utilize dot product of the latent embeddings of video and music to calculate the matching degree of a video-music pair.}

Considering that each video $\mathbf{v}$ consists of multimodal features (\textit{i.e.}, visual and textual features $\mathbf{v}_v$ and $\mathbf{v}_t$), both of which are important for the matching of background music, the generative process is extended to multimodal scenarios as shown in Figure \ref{FIG:MODEL_brief}, where the cross-generation is specified with finer granularity as follows:
\begin{align}
    &p(\mathbf{v}_v,\mathbf{v}_t \mid \mathbf{m})=\mathbb{E}_{\mathbf{z}_{m} \sim p\left(\mathbf{z}_{m} \mid \mathbf{m}\right)}\left[p\left(\mathbf{v}_{v} \mid \mathbf{z}_{m}\right)  p\left(\mathbf{v}_{t} \mid \mathbf{z}_{m}\right)\right] \\
    &p(\mathbf{m} \mid \mathbf{v}_v,\mathbf{v}_t)=\mathbb{E}_{\mathbf{z}_{v} \sim p\left(\mathbf{z}_{v} \mid \mathbf{v}_v,\mathbf{v}_t\right)}\left[p\left(\mathbf{m} \mid \mathbf{z}_{v}\right)\right],
\end{align}
{where the music latent embedding $\mathbf{z}_{m}$ is assumed to generate both visual and textual features of the video, and the video latent embedding $\mathbf{z}_{v}$ generates the matched music feature.}

To infer the joint video latent embedding $\mathbf{z}_v$ based on the complementary information from both visual and textual modalities, inspired by \cite{wu2018multimodal}, we assume the observations of the two modalities, $\mathbf{v}_{v}, \mathbf{v}_{t}$, are conditional independent given $\mathbf{z}_v$, where the joint posterior can be factorized into modality-specific posteriors as follows:
\begin{equation}
\label{eq:poe}
    p(\mathbf{z}_{v} \mid \mathbf{v}_v, \mathbf{v}_t) \propto  p(\mathbf{z}_{v_v}  \mid  \mathbf{v}_v) p(\mathbf{z}_{v_t} \mid \mathbf{v}_t),
\end{equation}
\noindent where $\mathbf{z}_v$ is the joint latent embedding inferred from multimodal video features. $\mathbf{z}_{v_v}$ and $\mathbf{z}_{v_t}$ are latent embeddings inferred from the visual modality and the textual modality, respectively. Eq. (\ref{eq:poe}) shows that the multimodal fusion of the bimodal information of micro-videos is in essence a product-of-experts (PoE) system. For Gaussian variables, the product is also Gaussian where the new mean and new variance become:
\begin{align}
    {\mathbf{\mu}_v}  &=\frac{ \mathbf{\mu} _ {v_v} \odot \left(1 / \mathbf{\sigma} _ {v_v} ^ {2}\right) + {\mathbf\mu} _ {v_t} \odot \left(1 / \mathbf{\sigma} _ {v_t} ^ {2}\right)}{  {1 / \mathbf{\sigma} _ {v_v} ^ {2}} + {1 / \mathbf{\sigma} _ {v_t} ^ {2}}} \label{eq:poe_1}\\
    {\mathbf\sigma_v} &= \operatorname{sqrt}\left(\frac{1}{  1 / \mathbf{\sigma} _ {v_v} ^ {2} + 1 / \mathbf{\sigma} _ {v_t} ^ {2}} \right) \label{eq:poe_2},
\end{align}
{where $\mathbf{\mu} _ {v_v}$ and $\mathbf{\mu} _ {v_t}$ are mean vectors of the visual and textual latent variables, $\mathbf{\sigma} _ {v_v}$ and $\mathbf{\sigma} _ {v_t}$ are variance vectors.}
Since theoretically, the mean of a Gaussian variable depicts its semantic structure and variance denotes uncertainty, the mean vector of the micro-video embedding is a weighted sum of the semantic information in each modality according to its informative level for the matching {purpose}. Therefore, the heterogeneous information from the visual and textual modalities is comprehensively fused where irrelevant information is down-weighted for a better generalization. Another by-product of such factorization is that, if the textual modality is missing in the training or test phase, which is a commonly encountered problem when the users are reluctant to add descriptions to their videos, we can temporarily drop the textual network in CMVAE and proceed the training or test process with the visual network.

After defining the inference process for the joint video variational posterior $\mathbf{z}_v$, our discussion shifts to the estimation of $\mathbf{z}_{v_v}$, $\mathbf{z}_{v_t}$ and $\mathbf{z}_m$ with the collected matching video-music pairs. Since the generative processes of the modalities and the matching status are parameterized as deep neural networks, the posterior distributions for $\mathbf{z}_{v_v}$, $\mathbf{z}_{v_t}$ and $\mathbf{z}_m$ are intractable. Therefore, we resort to variational inference, where we assume $\mathbf{z}_{v_v}$, $\mathbf{z}_{v_t}$ and $\mathbf{z}_m$ come from tractable families of distributions (which are also parameterized by deep neural networks) and in those families find the distributions closest to the true posteriors measured by the KL-divergence \cite{blei2017variational}. 
{Specifically, we approximate the true intractable posteriors with simpler variational posteriors, \textit{i.e.}, diagonal Gaussian distributions. Take the variational posterior of the music latent variable $\mathbf{z}_{m}$ for an example:
\begin{equation}
    q\left(\mathbf{z}_{m} \mid \mathbf{m}\right)=\mathcal{N}\left(\mathbf{\mu}_m, \operatorname{diag}\left\{\mathbf{\sigma}^{2}_m\right\}\right),
\end{equation}
where the variational posterior is parameterized by $\mathbf{\mu}_m$ and $\mathbf{\sigma}^{2}_m$. The parameters are outputs of the music inference neural network, which is also an MLP. The inference of variational posteriors of video visual and textual latent variables $\mathbf{z}_{v_v}$ and $\mathbf{z}_{v_t}$ is the same as the music latent variable with visual and textual inference networks. 
}

Previous work proves that the minimization of the KL-divergence {of variational posteriors and true posteriors} is equivalent to the maximization of the Evidence Lower BOund (ELBO) \cite{kingma2013auto}, which is composed of two parts: the cross reconstruction term $\mathcal{L}_{\text{cross\_recon }}$ and the KL-divergence term $\mathcal{L}_{\text{KL}}$. {The reconstruction part is as follows:}
\begin{equation}
\begin{aligned}
\mathcal{L}_{\text{cross\_recon }}&=\mathbb{E}_{\mathbf{z}_m \sim q\left(\mathbf{z}_m \mid \mathbf{m}\right)} [\log p(\mathbf{v}_v \mid \mathbf{z}_m)] \\
&+ \mathbb{E}_{\mathbf{z}_m \sim q\left(\mathbf{z}_m \mid \mathbf{m}\right)} [\log p(\mathbf{v}_t \mid \mathbf{z}_m)] \\
&+ \mathbb{E}_{\mathbf{z}_{v} \sim q\left(\mathbf{z}_v \mid \mathbf{v}\right)} [\log p(\mathbf{m} \mid \mathbf{z}_v)],
\end{aligned}
\label{eq:L_recon_loss}
\end{equation}
\noindent which aims to reconstruct the input features with latent variables. 
{The likelihood we consider in this paper is Gaussian, and therefore, maximizing the log-likelihoods for video $\mathbf{v}$ and music $\mathbf{m}$ conditioned on the music latent variable $\mathbf{z}_{m}$ and video latent variable $\mathbf{z}_{v}$ are equivalent to maximizing the negative mean square error (MSE) losses as:
\begin{equation}
\begin{aligned}
\mathcal{L}_{\text{cross\_recon }}&{=}-\sum\nolimits_{i} \left(\mathbf{v}_{v_ i}-f_{v_v}\left(\mathbf{z}_{m}\right)_{ i}\right)^{2} \\
 &-\sum\nolimits_{i} \left(\mathbf{v}_{t_i}-f_{v_t}\left(\mathbf{z}_{m}\right)_{ i}\right)^{2} \\
 &-\sum\nolimits_{i} \left(\mathbf{m}_{ i}-f_m\left(\mathbf{z}_{v}\right)_{ i}\right)^{2} \label{mse_2},
\end{aligned}  
\end{equation}
where $f_{v_v}$, $f_{v_t}$, and $f_m$ are non-linear functions implemented by video visual, video textual, and music generation networks, respectively. Through Eq. (\ref{mse_2}), the reconstruction term could be implemented by MSE losses between the observations and reconstructed inputs.}

The $\mathcal{L}_{{KL}}$ term penalizes the latent variables over deviating from the prior, which prevents encoding excessive information from the inputs and serves as a regularizer. It can be formulated as:
\begin{equation}
\begin{aligned}
\mathcal{L}_{\text{KL}} &=  \mathbb{E}_{\mathbf{z}_v \sim q(\mathbf{z}_v \mid \mathbf{v})}\left[D_{\text{KL}}\left(p\left(\mathbf{z}_{v} \mid \mathbf{v}\right) \ \mid  p\left(\mathbf{z}_{v}\right)\right)\right]  \\
&+ \mathbb{E}_{\mathbf{z}_m \sim q(\mathbf{z}_m \mid \mathbf{m})}\left[D_{\text{KL}}\left(p\left(\mathbf{z}_{m} \mid \mathbf{m}\right) \ \mid  p\left(\mathbf{z}_{m}\right)\right)\right], \\
\end{aligned}
\label{eq:kl}
\end{equation}
\noindent where $p\left(\mathbf{z}_{m}\right)$ and $p\left(\mathbf{z}_{v}\right)$ are the priors for the music and video latent variables, respectively, both of which are specified as the standard Normal distribution $\mathcal{N}(\mathbf{0}, \mathbf{I}_{d})$. {For Gaussian variables, the KL-divergence has an analytical solution.} In addition, inspired by \cite{bi_directional,wang2018learning}, the video-music matching loss $\mathcal{L}_{\text{matching}}$ that maximizes the $p(y \mid \mathbf{m},\mathbf{v})$ for matched pairs is specified as a bi-directional ranking loss as follows:
\begin{equation}
\label{eq:matching}
\begin{aligned}
\mathcal{L}_{\text{matching}}&= \sum_{{m}^+ ,m^-, v}\left[margin+{\mathbf{z}_{m^+}} \odot \mathbf{z}_{v}^{ \top}-{\mathbf{z}_{m^-}} \odot \mathbf{z}_{v}^{ \top}\right]_{+} \\
&+ \alpha \cdot \sum_{m ,v^+, v^-}\left[margin+{\mathbf{z}_{v^+}} \odot \mathbf{z}_{m}^{ \top} -{\mathbf{z}_{v^-}} \odot \mathbf{z}_{m}^{ \top}\right]_{+},\\
\end{aligned}
\end{equation}
\noindent where $[t]_{+}=\max (0, t)$. The first term of the right-hand side of Eq. (\ref{eq:matching}) is the video-to-music matching loss, where for a fixed video $v$, $m^+$ ,$m^-$ denote the matched music and unmatched music, respectively. {We constraint the distance between $\mathbf{z}_{v}$ and $\mathbf{z}_{m^+}$ to be smaller than the distance between $\mathbf{z}_{v}$ and $\mathbf{z}_{m^-}$, with a margin.}
The second term is the music-to-video matching loss, where the symbols are defined accordingly. 
{Through the bi-directional ranking loss, the distance of latent variables for a matched video-music pair is closer than that of unmatched pairs. Moreover, the relevant music clips are pairwise close in semantics since they are all relevant to the same video query, and the same to videos. Thus, the latent structures of the music and videos are well explored.
During training, the negative samples are selected in the mini-batch scale, where we select unmatched samples with top-$k$ inner-product similarities in the mini-batch as the negative samples for each term. In this way, the most hard-to-train negative samples are selected and the constraint that the distances between matched video-music pairs are smaller than unmatched ones could be maximized,  as well as the training efficiency is highly improved.}

Finally, combining the ELBO and matching loss, the joint training objective can be specified as:
\begin{equation}
    \mathcal{L}_{(\mathbf{z}_m, \mathbf{z}_v)} = \beta \cdot \mathcal{L}_{\text{cross\_recon }} - \mathcal{L}_{\text{KL}} + \gamma \cdot \mathcal{L}_{\text{matching}},
\end{equation}
\noindent where $\beta$ and $\gamma$ control the weight of the cross reconstruction loss and the matching loss, respectively. Note that to increase the model's robustness to modality missing in the test phase, we utilize a sub-sampled training strategy inspired by \cite{wu2018multimodal}, where the latent embedding of the visual modality $\mathbf{z}_{v_v}$ or the textual modality $\mathbf{z}_{v_t}$ of a micro-video is also constrained to complete the matching task by adding the single-modal objective {$\mathcal{L}_{(\mathbf{z}_m, \mathbf{z}_{v_v})}$ and $\mathcal{L}_{(\mathbf{z}_m, \mathbf{z}_{v_t})}$} to the joint training objective $\mathcal{L}_{(\mathbf{z}_m, \mathbf{z}_v)}$ as:
{
\begin{equation}
    \mathcal{L} = \mathcal{L}_{(\mathbf{z}_m, \mathbf{z}_v)}+\mathcal{L}_{(\mathbf{z}_m, \mathbf{z}_{v_v})} + \mathcal{L}_{(\mathbf{z}_m, \mathbf{z}_{v_t})}.
\end{equation}
Specifically, $\mathcal{L}_{(\mathbf{z}_m, \mathbf{z}_{v_v})}$ and $\mathcal{L}_{(\mathbf{z}_m, \mathbf{z}_{v_t})}$ replace the term of joint video latent variable $\mathbf{z}_v$ with visual and textual latent variables $\mathbf{z}_{v_v}$ and $\mathbf{z}_{v_t}$ in the training objective. By doing so,  our model could deal with modality missing in the test phase by using the remaining posteriors to approximate the joint posteriors.
During training, we use the reparameterization trick \cite{kingma2013auto} to maximize the above objective.
Mathematically, we sample $\epsilon \sim \mathcal{N}(\mathbf{0},  \mathbf{I}_d )$, and reparametrize the latent embedding
$\hat{\mathbf{z}}_{k}=\mathbf{\mu}_{k}+\mathbf{\sigma}_{k} \odot \epsilon$ with $k \in \{m, v, v_v, v_t\}$.
Thus, the stochasticity in the sampling process is isolated and the whole model can get the gradient when performing the back-propagation. This reparameterization operation can also be interpreted as a corruption where the latent mean vectors are injected with Gaussian noise and the latent variance vectors influence the noise level. In this way, more robust presentations could be obtained in the training phase.}
The detailed training steps of the proposed CMVAE are summarized in Algorithm \ref{alg:vae} for reference.

\begin{algorithm}[t]
\DontPrintSemicolon  \KwIn{A video-music matching dataset $\mathcal{D}=\{\mathcal{V}, \mathcal{M}, \mathrm{f}\}$;\\
 \quad \quad the video feature set $\mathcal{V}$ contains the visual modality $\mathbf{v}_v$ and the textual modality $\mathbf{v}_t$ for each video $\mathbf{v}$;\\
 \quad \quad the music feature set $\mathcal{M}$ contains the audio modality $\mathbf{m}$;\\
 \quad \quad the mapping $\mathrm{f} = \{(\mathbf{v}, \mathbf{m}, y)\}$ with $y \in \{0, 1\}$;\\
 \quad \quad $\Theta$ indicates the model parameters.}

Randomly initialize $\Theta$.\;
 \While{not converged}{
  Randomly sample a batch $\hat{\mathcal{D}}$ from $\mathcal{D}$.\;
  \ForAll{$mod \in \{m, v_v, v_t, \}$}{
    Compute $\mu_{mod}, \sigma_{mod}$ via variational encoders.\;
  }
  
  Compute $\mathbf{\mu}_{v}, \mathbf{\sigma}_{v}$ of $p(\mathbf{z}_{v} \mid \mathbf{v}_v, \mathbf{v}_t)$ from $(\mathbf{\mu}_{v_v}, \mathbf{\sigma}_{v_v})$ and $(\mathbf{\mu}_{v_t}, \mathbf{\sigma}_{v_t})$ with the PoE module {via Eq. (\ref{eq:poe_1}) and Eq. (\ref{eq:poe_2}).}\;
  Add the KL-divergence {via Eq. (\ref{eq:kl})} to the loss.\;
  Sample $\mathbf{\epsilon} \sim \mathcal{N}\left(\mathbf{0}, \mathbf{I}\right)$ and compute $\mathbf{z}_m$ and $\mathbf{z}_v$ via the reparametrization trick.\;
  Add the cross reconstruction loss and matching loss as Eq. (\ref{mse_2}) and Eq. (\ref{eq:matching}) to  $\mathcal{L}_{(\mathbf{z}_m, \mathbf{z}_v)}$.\;
  Add the loss of modality-specific $\mathcal{L}_{(\mathbf{z}_m, \mathbf{z}_{v_v})}$ and $\mathcal{L}_{(\mathbf{z}_m, \mathbf{z}_{v_t})}$ to get final $\mathcal{L}$.\;
  Compute the gradient of loss $\nabla_{\Theta} \mathcal{L}$.\;
  
  Update $\Theta$ by taking stochastic gradient steps.\;
}
\Return{$\Theta$}\;
\KwOut{CMVAE model trained on dataset $\mathcal{D}$.}
\caption{{CMVAE-SGD:} Training CMVAE with SGD.}
\label{alg:vae}
\end{algorithm}

\section{EXPERIMENTS} \label{section5}
In this section, we conduct extensive experiments based on the established TT-150k dataset to evaluate our proposed model CMVAE. Specifically, we aim to answer the following research questions:
\begin{itemize}[leftmargin=*]
\item How does CMVAE perform compared with the baseline methods for the micro-video background music recommendation? {Among them, content-based and variational-based cross-modal matching methods are included for comprehensive comparisons.}
\item {Since the cross-generation strategy
and the PoE fusion module are the main designs in our proposed CMVAE, we investigate how do the two components contribute to the performance as ablation studies.}
\item {Textual modality missing is a commonly encountered problem when users are reluctant to add descriptions to their videos. Therefore, we conduct experiments to verify the adaptation of CMVAE to the realistic background music recommendation scene by checking its performance when textual modality is missing in the test phase.}
\end{itemize}

{Moreover, a visualization of the recommendation results by CMVAE is included to qualitatively assess the performance.}

\begin{table*}[htbp]
\centering
\caption{Comparison between the proposed CMVAE and various baselines under weak and strong generalization scenarios.}
\label{TAB:result_with_different_methods}
\setlength{\tabcolsep}{3mm}
\begin{tabular}{l|cccc|cccc}
\toprule
&\multicolumn{4}{c|}{\textbf{Existing music (Weak generalization)}} & \multicolumn{4}{c}{\textbf{New music (Strong generalization)}}\\
&Recall@10 & Recall@15 & Recall@20 &Recall@25& Recall@10 & Recall@15 & Recall@20 &Recall@25 \\ 
\midrule
CMVAE& \textbf{0.2606} & \textbf{0.3684} & \textbf{0.4489} & \textbf{0.5137} & \textbf{0.1610} & \textbf{0.2274} & \textbf{0.2881} & \textbf{0.3479} 
\\
\midrule
Random & 	\underline{0.0959}&		\underline{0.1398}&  \underline{0.1788}&	\underline{0.2191}& \underline{0.0978}&  \underline{0.1443}& 	\underline{0.1866}& \underline{0.2268}
 \\
PopularRank & 	0.0689 & 0.1030&	0.1390&	 0.1766 & - &-&-&-
 \\
\midrule
CCA \cite{hardoon2004canonical} & 0.0465& 0.0640& 0.0934& 0.1278& 0.0412& 0.0670& 0.0941& 0.1273

\\
CEVAR \cite{suris2018cross}& \underline{0.1262} & \underline{0.1749} & \underline{0.2302} & \underline{0.2794} & \underline{0.1223} & \underline{0.1823} & \underline{0.2416} & \underline{0.3019} 
\\
CMVBR \cite{hong2018cbvmr}& 0.0927& 0.1431& 0.1929& 0.2425& 0.0966& 0.1441& 0.1940& 0.2395
\\
DSCMR \cite{zhen2019deep}&  0.1062& 0.1553& 0.2077& 0.2651& 0.1133& 0.1775& 0.2241&  0.2793
\\
UWML \cite{wei2020universal}& 0.1053& 0.1554& 0.2103& 0.2573& 0.0933& 0.1299& 0.1729& 0.2255\\
\midrule
Dual-VAE \cite{shen2018deconvolutional}& 0.1607& 0.2272& 0.2897& 0.3514& \underline{0.1450}& \underline{0.2143}& {0.2705}& \underline{0.3294}
\\
DA-VAE \cite{davae}& 0.1645& 0.2390& 0.3063& 0.3698&  0.1417& 0.2104&  \underline{0.2740}& 0.3280
\\
Cross-VAE \cite{yu-etal-2020-crossing}& \underline{0.2584} & \underline{0.3590} & \underline{0.4360} & \underline{0.5077} & {0.1391} & {0.2008} & {0.2706} & {0.3233} 
\\
\bottomrule
\end{tabular}
\end{table*}

\subsection{Methods for Comparisons}
To demonstrate the effectiveness of the proposed CMVAE, we draw comparisons with several state-of-the-art cross-modal matching baselines. For background music recommendation in this article, each video has exactly one background music as the ground truth, so the co-occurrence information necessary for collaborative filtering methods does not exist. Therefore, collaborative-based methods (\textit{e.g.} NCF \cite{he2017neural}) are not applicable to our paper. For methods where the original task is to match two sources with a single modality, we concatenate the visual and textual features as the ``single modality" representations of the micro-videos. The baselines are listed as follows:

\begin{itemize}[leftmargin=*]
    \item \textbf{Random}: For each video, $K$ music clips are randomly selected as the candidates for recommendation.
    \item \textbf{PopularRank}: The number of adoption of a music clip, \textit{i.e.,} its popularity, is utilized to select and recommend top-$K$ popular music clips for every video.
    \item \textbf{CCA} \cite{hardoon2004canonical}: CCA (Canonical correlation analysis) is a widely adopted multivariate correlation analysis method. We extract two canonical variates from music features and video features via CCA, and then calculate cosine similarity for matching. 
    \item \textbf{CEVAR} \cite{suris2018cross}: {CEVAR aims to make the visual and audio embeddings which come from the same video close using a cosine similarity loss with the supervision loss of the corresponding class of the video optimized together. To make it amenable to our problem, we eliminate the label prediction loss for comparison.} 
    \item \textbf{CMVBR} \cite{hong2018cbvmr}: CMVBR is a content-based method that introduces a soft intra-modal structure loss to better align intra-modal features for matching two multimodal sources.
    \item \textbf{DSCMR} \cite{zhen2019deep}: DSCMR is a supervised model with {semantic category labels of the samples attached}, which minimizes the recognition loss in the label space and the common representation space. A weight-sharing strategy is used to reduce the cross-modal difference of multimedia data. We eliminate the recognition loss of the label space to make it applicable for our dataset.  
    \item \textbf{UWML} \cite{wei2020universal}: UWML is a metric learning-based method with a triplet loss utilized to encourage the closeness of positive pairs than that of negative pairs. It introduces a weighting framework for the positive and negative pairs where a larger weight calculated by polynomial functions based on the similarity scores is assigned to a harder-to-match pair. 
\end{itemize}

We also compare our proposed cross-modal variational auto-encoder with the state-of-the-art variational-based generative models:

\begin{itemize}[leftmargin=*]
    \item \textbf{Dual-VAE} \cite{shen2018deconvolutional}: Dual-VAE aims to exploit deconvolution on word sequence in the decoder part in a text-based retrieval model. It is trained jointly on question-to-question and answer-to-answer reconstructions. 
    \item \textbf{DA-VAE} \cite{davae}: This dual-aligned variational auto-encoder focuses on the situation where samples in some modalities are missing for image-text retrieval. DA-VAE utilizes an entropy-maximization constraint to further align two modalities along with the self-reconstruction loss.
    \item \textbf{Cross-VAE} \cite{yu-etal-2020-crossing}: This crossing variational auto-encoder utilizes a cross-reconstruction strategy where answers are reconstructed from question embeddings and questions are reconstructed from answer embeddings for text-based retrieval. 
\end{itemize}

For the above three methods, we only use the corresponding cross-modal generation module to replace the PoE-based multimodal cross-generation module in the proposed CMVAE for a fair comparison, where we concatenate the multimodal features of the micro-video to be the query modality. 
Dual-VAE and DA-VAE both optimize the self-reconstruction loss and the matching loss, whereas DA-VAE further utilizes an entropy-maximization constraint to better align two modalities. For Cross-VAE, a cross-reconstruction loss is utilized along with the matching loss to form the training objective.

\subsection{Evaluation Metric}
We use Recall@$K$ \cite{wei2020universal} to evaluate the model performance, which is a standard metric for cross-modal retrieval that calculates the average hit ratio of the matched music clips over all the music clips that are ranked with top $K$ match scores. Since in our dataset, the popularity of different music clips varies considerably, as shown in Figure \ref{FIG:sampling}, weighting the match equally for different music clips could lead to systematic bias that favors the popular music clips. Therefore, following \cite{ji2019recommendation}, music clips for evaluation are weighted by the inverse of their popularity levels (\textit{i.e.,} estimated propensity scores), and thus the Recall@$K$ we adopt in this paper is formalized as follows:
\begin{align}
    \text{Recall}@K &=\sum_{v \in \mathcal{V}^{t e} } \lambda_v \cdot \#\text {Hits }_{v} @ K \label{eq:recall}\\
    \lambda_v &= \frac{1 / P(v \rightarrow m)}{\sum_{i \in \mathcal{V}^{t e} } 1/P(i \rightarrow m)} \label{eq:weight},
\end{align}
where $\mathcal{V}^{t e}$ represents the set of test videos and $\lambda_v$ denotes the weight of the video $v$ calculated by Eq. (\ref{eq:weight}). Specifically, $1 / P(v \rightarrow m)$ represents the reciprocal of the popularity of the music clip that the target video $v$ used as background music. With Eq. (\ref{eq:recall}), the $\text{Recall} @K $ used in this paper calculates the popularity-debiased hit-ratio of relevant music clips ranked among the top-$K$ recommendation list for all videos in the test set.

\subsection{Experimental Setup}
{
We use two dataset split strategies to create train, validation, and test sets. The first strategy is the ``weak generation”, where for each background music clip, the interacted videos are split by a ratio of 8:1:1 to construct train, validation, and test sets. After splitting the dataset, we have 16,036 video-music pairs for the test set. In this case, although the videos in the test set are new, the candidate music clips in the test set have been adopted by at least one micro-video in the training set. }

{ The second strategy we consider is ``strong generalization” where the music clips in the test set are not present in the training set. To keep the popularity distribution of music candidates in the test set identical to that of the training set, a stratified sampling strategy is adopted to split the music clips. Specifically, in each of the popularity stratum, we randomly sample the music clips with an 8:1:1 ratio for train, validation, and test sets with all their associated videos. After splitting the dataset, 14,656 video-music pairs are sampled for the test set. The ``strong generalization” is relatively more difficult than "weak generalization" where the music clips do not appear on the training set and are not trained for evaluations. We consider it to be more realistic and robust as well. By default, we evaluate the models under both the weak and strong generalization scenarios. For weak generalization, all the music clips in the test set have been adopted by at least one micro-video in the training set, while for strong generalization, music clips in the test set are not present in the training stage.  We report evaluation scores averaged over three different randomly sampled train/validation/test splits. }

Our CMVAE model is implemented in Pytorch. The embedding size is fixed to 512 and the batch size is set to 1,024 for all models, empirically. {The overall architecture for the music and video networks would be $[F \rightarrow 64 \rightarrow 64 \rightarrow 64 \rightarrow F]$ with $F$ to be the dimension of different features.} We use ReLU as the non-linear activation. For all the models, the hyperparameters are selected based on evaluation metrics on the validation set. The learning rate is tuned amongst \{0.0001, 0.005, 0.001, 0.05, 0.01\}. By searching, we set the weight of the L2 norm penalty as 0.001, the weight of the music-to-video matching loss $\alpha$ as 3, and choose the ten most confusing negative samples in the bi-directional ranking loss during training. For strong and weak generalizations, we set the weight of the reconstruction loss $\beta$ to $1e^{5}$ and the learning rate to 0.05 and 0.0001, respectively.

\subsection{Performance Comparison with Baselines}

The comparisons between the proposed CMVAE and the state-of-the-art baselines are summarized in Table \ref{TAB:result_with_different_methods}. For strong generalization, since the music clips in the test set are not present at the training stage, we assume that the popularity of the test music clips is unknown or cannot be accurately estimated due to the lack of data, and thus exclude the PopularRank method from comparisons. 

Among the methods that we draw comparisons with, CCA finds a linear projection that maximizes the correlation between projected vectors of two different modalities. The inferior performance of CCA demonstrates that the matching pattern between video and music could not be modeled by a simple linear relation. For the deep learning-based methods, CEVAR and CMVBR are originally designed for video-music matching, whereas DSCMR and UWML are designed for matching between visual (\textit{i.e.}, image or video) and textual modalities. Therefore, the modules that are specifically designed for image-text matching tasks such as stacked cross attention network \cite{Lee_2018_ECCV}, which aligns image regions and text words, are not applicable to our task and are therefore removed. We find that directly extending these image-text matching methods to the video-music setting where one of the sources is composed of multiple modalities could not achieve satisfying results. The reason may be that the patterns that are responsible for the matching between videos and background music are more elusive than that of image-text matching, and therefore the naive matching loss could not force the model to capture the complex matching patterns for the alignment of the music and video latent spaces. CEVAR performs the best among the selected deep learning-based baselines. Compared to CMVBR, CEVAR utilizes a cosine matching loss to constrain the closeness of the embeddings of the matched video-music pairs instead of the inner-product loss, which eliminates the systematic difference of image embeddings due to varied lightness or saturation, \textit{etc.} 

The generative-based matching approaches listed at the bottom of Table \ref{TAB:result_with_different_methods} improve significantly over the matching-loss-based baselines. Among them, Dual-VAE and DA-VAE employ self-reconstruction, which leads to a more structured shared latent space for the micro-video and music embeddings. DA-VAE further imposes an entropy-maximization constraint on the joint representations motivated by Jaynes’s theory \cite{PhysRev.108.171} which improves the performance over Dual-VAE. {By replacing the self-reconstruction module with a cross-reconstruction module compared to Dual-VAE}, Cross-VAE performs significantly better than DA-VAE for weak generalization and performs on par with DA-VAE for strong generalization.

CMVAE performs significantly better compared to all other methods. We attribute the improvement of CMVAE to the following two designs: 1) CMVAE takes advantage of both generative-based and matching-loss-based models, which is optimized against the composite loss consisting of the cross reconstruction loss, the bi-directional ranking loss, and the VAE regularization loss. Therefore, the latent embeddings are constrained to encode more matching-relevant information from the videos and music while regularizing them from overfitting. 2) By aggregating information in the textual and visual modalities with the PoE module, the information in each modality is fused according to its importance to the matching purpose. Consequently, the irrelevant and redundant information contained in the micro-video embeddings is reduced such that a more robust generalization can be achieved for the model.

\begin{table}[t]
\centering
\caption{The results of Recalls on different generation strategies.}
\begin{subtable}[t]{\columnwidth}
\centering
\caption{\textbf{Existing music}}\label{tab:result_with_cross}

\begin{tabular}{lcccc}
\toprule                                     
{}& {Recall@10} & {Recall@15} & {Recall@20} & {Recall@25} \\
\midrule
$\text{CMVAE}_\text{w/o}$  & 0.1305& 0.1979& 0.2583& 0.3173                     \\
$\text{CMVAE}_\text{Dual}$ & 0.1637& 0.2302& 0.2937& 0.3544
 \\  
$\text{CMVAE}_\text{Cross}$& \textbf{0.2606}& \textbf{0.3684}& \textbf{0.4489}& \textbf{0.5137}
\\
\bottomrule
\end{tabular}
\end{subtable}

\bigskip

\begin{subtable}[t]{\columnwidth}
\centering
\caption{\textbf{New music}}\label{tab:result_with_cross_new}
\begin{tabular}{lcccc}
\toprule      
{}& {Recall@10} & {Recall@15} & {Recall@20} & {Recall@25} \\
\midrule
{$\text{CMVAE}_\text{w/o}$} & 0.1163& 0.1767& 0.2287& 0.2935  \\
{$\text{CMVAE}_\text{Dual}$} &  0.1470& 0.2183& 0.2730& 0.3314 \\
{$\text{CMVAE}_\text{Cross}$} & \textbf{0.1610}& \textbf{0.2274}& \textbf{0.2881}& \textbf{0.3479}  \\
\bottomrule
\end{tabular}
\end{subtable}

\label{TAB:result_with_cross}
\end{table}

\subsection{Ablation Study of CMVAE}
In this section, we investigate the effectiveness of different components in CMVAE. In detail, as the cross-generation strategy plays a vital role in CMVAE, we compare it with other generation strategies. Moreover, we explore the effectiveness of PoE fusion by comparing it with other variational multimodal fusion methods. 

\subsubsection{The effectiveness of cross-generation strategy}

To explore the effectiveness of the cross-generation strategy which is a core component of our method, we compare different generation strategies with cross-generation. The method with only matching loss and without the generation module \cite{yang2019multilingual} and the method with the dual-generation (self-reconstruction strategy with VAE is utilized) \cite{shen2018deconvolutional} are selected for comparisons. We can see significant performance improvements with results shown in Table \ref{TAB:result_with_cross}. Superior results verify that the cross-generation strategy can force the latent embeddings of videos and background music clips to align with each other where similarities could be calculated to recommend music clips to videos. However, the dual-generation strategy can only enhance the modeling of latent embeddings by constraining the video latent embeddings to generate video features and the music latent embeddings to generate music features. While the alignment of video and music embeddings is not guaranteed. Inferior results yield on the method without using any generation strategy where the semantic-rich video latent space and the monotonous music latent space are naturally hard to align for matching videos and music. {Moreover, the weak association between videos and background music makes it hard to learn the matching patterns with the matching loss and perform effective recommendations.}

\subsubsection{The effectiveness of PoE fusion}

\begin{table}[]
\caption{The results of Recalls on different multimodal variational models for videos.}
\begin{subtable}[t]{\columnwidth}
\centering
\caption{\textbf{Existing music}}\label{tab:result_with_MoE}

\begin{tabular}{lcccc}
\toprule                                     
{}& {Recall@10} & {Recall@15} & {Recall@20} & {Recall@25} \\
\midrule
{$\text{CMVAE}_\text{JMVAE}$} & 0.1990& 0.2929& 0.3673& 0.4453    \\
{$\text{CMVAE}_\text{MoE}$} & 0.2144&  0.3058&  0.3856& 0.4498 \\
{$\text{CMVAE}_\text{PoE}$} & \textbf{0.2606}& \textbf{0.3684}& \textbf{0.4489}& \textbf{0.5137}  \\
\bottomrule
\end{tabular}
\end{subtable}

\bigskip

\begin{subtable}[t]{\columnwidth}
\centering
\caption{\textbf{New music}}\label{tab:result_with_modal_missing_new}
\begin{tabular}{lcccc}
\toprule      
{}& {Recall@10} & {Recall@15} & {Recall@20} & {Recall@25} \\
\midrule
{$\text{CMVAE}_\text{JMVAE}$} & 0.1408&  0.2008&  0.2633& 0.3325  \\
{$\text{CMVAE}_\text{MoE}$} &  0.1238&  0.1879&  0.2614& 0.3249 \\
{$\text{CMVAE}_\text{PoE}$} & \textbf{0.1610}&  \textbf{0.2274}& \textbf{0.2881}& \textbf{0.3479}  \\
\bottomrule
\end{tabular}
\end{subtable}

\label{TAB:result_with_MoE}
\end{table}

In this section, we explore the effectiveness of PoE by comparing it with other multimodal variational fusion methods:  the mixture-of-experts (MoE) \cite{shi2019variational} which assumes the joint variational posterior follows a Gaussian mixture distribution, and the joint variational auto-encoder (JMVAE) \cite{suzuki2016joint} which concatenates encoded features of multiple modalities to learn a joint variational posterior. Table \ref{TAB:result_with_MoE} shows that CMVAE with the PoE module performs significantly better than the other two methods for both the strong and weak generalizations. The superior results demonstrate that the video latent variable, where the mean is a sum of mean vectors from visual and textual modalities weighted by the reciprocal of corresponding variance, contains less irreverent information and therefore improves the model generalization. In contrast, MoE treats each modality as equally important and JMVAE simply concatenates the available modalities as the micro-video representations, which makes the utilization of multimodal information less effective than the PoE module used in CMVAE.

\subsection{Robustness to Modality Missing}

\begin{table}[]
\caption{The results of Recalls on modality missing problem in the test phase. (V: Visual, T: Textual)}

\begin{subtable}[t]{\columnwidth}
\centering
\caption{\textbf{Existing music}}\label{tab:result_with_modal_missing_exsiting}

\begin{tabular}{lcccc}
\toprule                                     
{}& {Recall@10} & {Recall@15} & {Recall@20} & {Recall@25} \\
\midrule
{V} & 0.2479  & 0.3472  & 0.4281& 0.4977  \\
{V + T} & 0.2606  & 0.3684  & 0.4489& 0.5137  \\
\bottomrule
\end{tabular}
\end{subtable}

\bigskip

\begin{subtable}[t]{\columnwidth}
\centering
\caption{\textbf{New music}}\label{tab:result_with_modal_missing_new}
\begin{tabular}{lcccc}
\toprule      
{}& {Recall@10} & {Recall@15} & {Recall@20} & {Recall@25} \\
\midrule
{V} & 0.1573  & 0.2275  & 0.2849 & 0.3459  \\
{V + T} & 0.1610  & 0.2274  & 0.2881& 0.3479  \\
\bottomrule
\end{tabular}
\end{subtable}

\label{tab:result_with_modal_missing}
\end{table}

In the real-world scenario, it is inevitable to encounter modality missing when analyzing micro-video contents. For example, some casual uploaders may be reluctant to write a title or hashtags when posting a video. Therefore, it is crucial for our multimodal model to be able to deal with the modality missing problem in the test phase without training multiple networks corresponding to each subset of modality combinations. With the PoE module to fuse the multimodal information, our model can easily address the modality missing problem by using the remaining visual embedding as a surrogate to the joint video embedding. To explore the performance of CMVAE when modality missing occurs at the test phase, we eliminate the textual modality of micro-videos in the test sets and report the performance. Results listed in Table \ref{tab:result_with_modal_missing} show that the performance decrease of CMVAE due to the missing of textual modality is minor. Since CMVAE is trained on the sub-sampled training paradigm \cite{wu2018multimodal} where the latent embeddings of a single modality are forced to do the matching task, such results demonstrate the robustness of the textual modality missing problem.

\subsection{Qualitative Assessment}
\begin{figure*}
\centering
\includegraphics[scale=1.8]{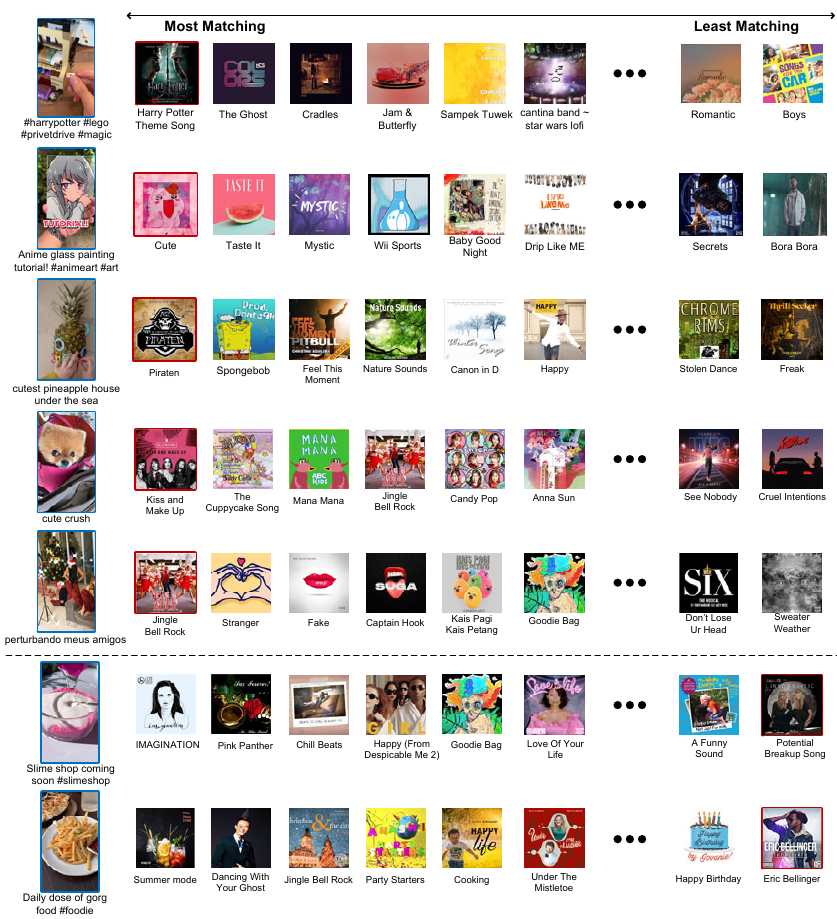}
\caption{Qualitative assessment of CMVAE for micro-video background music recommendation by visualizing some examples of query videos for the test set and the retrieved music clips ranked by their matching levels. The pictures in blue on the left are the representative frames of the micro-videos. The pictures on the right are the cover images of the music clips which are ranked by their matching level to the micro-video judged by CMVAE with the ground-truth music clips marked in red.}
\label{FIG:visualization}
\end{figure*}

We conduct experiments to visualize some examples of query videos for the test set and the retrieved music clips ranked by their matching levels as illustrated in Figure \ref{FIG:visualization} to better check the effectiveness of the recommendations of our method. Specifically, the pictures in blue on the left are the representative frames of the micro-videos with the textual information attached. The pictures on the right are the cover images of the music clips which are ranked by their matching level to the micro-video judged by CMVAE with the ground-truth music clips marked in red.

From the figure, we can see that our CMVAE can recommend suitable background music clips by aligning video and music latent embeddings via cross-generation using multi-modal information. Moreover, the expression and atmosphere the video convey contained in the visual and textual information are matched to the atmosphere expressed in the background music recommended by CMVAE. Take the first two given videos in the figure as examples, the uppermost video expresses the atmosphere of magic and the music ranked top few basically is in ethereal tunes. The video in the second line shows a painting procedure that matches more lovely and light music by our model. 
{At the same time, we have visualized two bad cases which our model failed to recommend. From the last two lines of examples, we could see that when users do not consider the matching degree of the music with the video content but choose the most popular music while selecting the background music, our model may fail to recommend the ground truth. Since our proposed CMVAE does not model users' preference bias for music popularity, it is difficult to recommend the popular-but-unmatched ground truth.  However, from another aspect, this also inhibits the Matthew effect of music popularity on recommendation, which prevents popular music from being overly recommended while the truly matched music is ignored.  We refer readers to some existing work on popularity debias \cite{popular_bias,intervention_debias,model_agnostic}.}

\section{CONCLUSIONS}  \label{section6}

In this paper, we introduce CMVAE, a hierarchical Bayesian cross-modal generative model for content-based micro-video background music recommendation. To solve the problem of lacking a publicly available dataset, we establish a large-scale database, TT-150k, which contains extracted features from more than 3,000 candidate music clips and about 150k micro-videos with the popularity distribution reflecting the real-world scenario. Experimental results demonstrate that by modeling the matching of relevant music to micro-videos as a multimodal cross-generation problem with PoE as the fusion strategy, CMVAE can significantly improve the background music recommendation quality compared to state-of-the-art methods, and it is robust to textual modality missing problem in the test phase.

\ifCLASSOPTIONcaptionsoff
  \newpage
\fi

\bibliographystyle{IEEEtranN}
\bibliography{refs}

\begin{IEEEbiographynophoto}{Jing Yi}
is pursuing the Ph.D. degree with Wuhan University, China. 
\end{IEEEbiographynophoto}

\begin{IEEEbiographynophoto}{Yaochen Zhu}
is pursuing the M.Eng. degree with Wuhan University, China.
\end{IEEEbiographynophoto}

\begin{IEEEbiographynophoto}{Jiayi Xie}
is pursuing the Ph.D. degree with Wuhan University, China. 
\end{IEEEbiographynophoto}

\begin{IEEEbiographynophoto}{Zhenzhong Chen}
is a Professor with Wuhan University, China. 
\end{IEEEbiographynophoto}

\end{document}